# MHAttnSurv: Multi-Head Attention for Survival Prediction Using Whole-Slide Pathology Images

Shuai Jiang, MHS[a], Arief A. Suriawinata, MD[b], Saeed Hassanpour, PhD[a,c,d*]

[a]Department of Biomedical Data Science, Geisel School of Medicine at Dartmouth, Hanover, NH 03755, USA

[b]Department of Pathology and Laboratory Medicine, Dartmouth-Hitchcock Medical Center, Lebanon, NH 03756, USA

[c]Department of Computer Science, Dartmouth College, Hanover, NH 03755, USA

[d]Department of Epidemiology, Geisel School of Medicine at Dartmouth, Hanover, NH 03755, USA

[*] Corresponding Author: Saeed Hassanpour, PhD

Postal address: One Medical Center Drive, HB 7261, Lebanon, NH 03756, USA

Telephone: (603) 650-1983

Email: Saeed.Hassanpour@dartmouth.edu

# Abstract

In pathology, whole-slide images (WSI) based survival prediction has attracted increasing interest. However, given the large size of WSIs and the lack of pathologist annotations, extracting the prognostic information from WSIs remains a challenging task. Previous studies have used multiple instance learning approaches to combine the information from multiple randomly sampled patches, but different visual patterns may contribute differently to prognosis prediction. In this study, we developed a multi-head attention approach to focus on various parts of a tumor slide, for more comprehensive information extraction from WSIs. We evaluated our approach on four cancer types from The Cancer Genome Atlas database. Our model achieved an average c-index of 0.640, outperforming two existing state-of-the-art approaches for WSI-based survival prediction, which have an average c-index of 0.603 and 0.619 on these datasets. Visualization of our attention maps reveals each attention head focuses synergistically on different morphological patterns.

# 1. Introduction

Each year, nearly 2 million people living in the United States are diagnosed with cancer; one-third of these patients will die within five years (American Cancer Society, 2021). The accurate and timely prediction of patient survival is crucial for shared clinical decision-making, treatment planning, and patient psychological adjustments (Hagerty et al., 2005). While both host- and environment-related factors can affect the survival of the cancer patient, the most fundamental prognostic factor is the tumor itself (Gospodarowicz and O'Sullivan, 2003). Pathological examination is a routine procedure at the time of diagnosis to determine the type and malignancy of tumors. In addition to its utility in diagnosis, the histopathological features of cancer, including tumor size, lymph node involvement, and metastasis, are commonly used in survival prediction models (Feng et al., 2019; Phung et al., 2019), proving the prognostic value of morphological features of tumors. However, the massive amount of information in one histology slide presents formidable challenges. Manual information extraction from the microscopic examination is time-consuming, and the amount of information structurally collected in pathology reports is often limited (Betge et al., 2012; Morris et al., 2007).

Whole-slide images or WSIs are the digitized versions of the complete microscope slides scanned at high resolution. There has been a growing body of work on the application of deep learning methods for predicting prognosis using WSIs with mixed results (Cheerla and Gevaert, 2019; Mobadersany et al., 2018; Wulczyn et al., 2020; Yao et al., 2019; Zhu et al., 2017b). A typical WSI can occupy several gigabytes of storage, while the lesions are usually small. Considering all regions with equal importance could miss the areas that are critical for prognosis. It has been reported that incorporating region-of-interest (ROI) annotations on WSIs could improve prognosis prediction performance (Wulczyn et al., 2021); however, such annotations require extensive time and expertise and are usually not readily available. The lack of ROI annotations on WSIs limits the application of machine learning prognostic tools in clinical practice. Besides, there are several other challenges in using WSIs for prognosis prediction: 1) the WSIs are too large, consequently most deep learning models are either computationally infeasible or methodologically incapable of handling them; 2) accurate survival prediction requires both the holistic representation of a WSI to evaluate the extent of the tumor and the detailed features at the cellular level to assess tumor characteristic, which are the opposite ends of the spectrum in image analysis; and 3) the whole slide datasets with survival information are usually of limited sample size, and the outcome information is only available for part of the participants either due to the study ends before the event occurs or the participants leave during follow-up (i.e., right censoring). The insufficient number of events makes a deep learning model overfit easily and the evaluation of the model performance dependent on data partitions.

## 1.1. Related work

### 1.1.1 Manually generated features

Early studies have relied on manually extracted features for survival analysis. Tumor size, grade, and stage are commonly used in statistical models for prognosis prediction (Liu et al., 2017; Rosenberg et al., 2005). These features are routinely collected by pathologists and are available from pathology reports. Despite the high prognostic values of these variables, these crude partitions cannot provide further risk stratifications. More specific biomarkers, such as mitotic index (MI), have been shown to be an independent predictor of long-term survival of cancer patients (Van Diest and Van Der Wall, 2004). While these biomarkers usually have strong clinical significance, obtaining them requires extensive inputs from domain experts. This process can be time-consuming and subject to human discretion. Moreover, the types of prognostic pathological factors might only be specific to certain cancer types. Another related method is to use high-throughput cell image analysis tools (e.g., CellProfiler) to extract predefined cell features automatically, such as size, shape, intensity, and texture (Carpenter et al., 2006; Chen et al., 2021). However, these handcrafted features are usually limited in nature and relatively redundant (Caicedo et al., 2017), and may not contain high-level prognostic information.

### 1.1.2 Automatic feature extraction with ROI annotation

With the evolution of computer vision techniques, automatic feature extraction using the latest deep learning models has become the standard practice. Nevertheless, processing an entire WSI is not a computationally feasible task given the extremely large size of WSIs. As a result, WSIs are broken into smaller images, called patches, with standard image sizes for typical computer vision tasks. This approach treats the survival prediction problem as a multiple instance learning (MIL) problem and does not consider the sequence or spatial location of the patches. As different patches usually contain redundant information, a sampling approach is used to select a subset of patches for model training and evaluation. The DeepConvSurv model developed by Zhu et al. sampled patches of size 1024×1024 from the ROIs of WSIs (Zhu et al., 2017a). These sampled patches were processed with a convolutional neural network (CNN) model so the local features within the patches could be extracted for survival prediction. Another method by Mobadersany et al. sampled one patch from the ROI region as the input for a CNN model (Mobadersany et al., 2018). During prediction time, multiple patches were sampled from each ROI region, and the median risk score was computed. For patients with multiple ROI regions, the second largest risk score was used as the final risk score for the patient. But these methods still rely on the expensive and subjective ROI annotations from pathologists, which limits their future applications. Furthermore, the prognostic information of an ROI can be incomplete, as morphological features outside of the tumor region can still be important for predicting patient risk.

*1.1.3 Automatic feature extraction without ROI annotation*

Recent studies have focused on predicting the survival status of patients without ROI annotation. Wulczyn et al. used a weakly supervised approach to sample several patches (e.g., 16 patches per case/iteration) randomly and feed them to a CNN model (Wulczyn et al., 2020). Average pooling was used to aggregate the feature vectors from multiple patches, and the resulted global feature vector was used for survival analysis. This full system (i.e., feature extraction + survival prediction) was trained end-to-end to obtain optimal results. The underlying assumption is that when a sufficiently large number of patches are sampled, the chance that at least one of them is discriminative will be large. This method was also used in a prior study for predicting prognosis and genetic profiling of lower-grade gliomas and has achieved satisfying results (Jiang et al., 2021). But because this method does not distinguish discriminative patches from non-discriminative ones, there is still room for further improvement.

In order to select the discriminative patches for survival prediction, the WSISA model resizes the patches into smaller thumbnails and uses K-means clustering to group them into several clusters (Zhu et al., 2017b). They then fit separate DeepConvSurv models for each cluster and select the clusters with high prognosis accuracy. The features from patches in the selected clusters are then aggregated for survival prediction. Despite being able to identify discriminative patch clusters, this method consists of several steps and needs to train multiple separate models. As a result, the interplay among different feature types cannot be captured and its performance is limited.

In another study, a separate model was trained to predict whether each patch belongs to ROI or not (Saillard et al., 2020). The predicted ROI probability was used as the weights to average patches. During inference time, the pre-trained ROI differentiation model predicts the probability that each patch belongs to the ROI. This approach only requires the ROI annotation at the training stage but not at the inference stage. A similar approach was used in another study but with hard segmentation instead of probability weighting (Wulczyn et al., 2021). Other studies experimented with graph models (Li et al., 2018) and Capsule Networks (Tang et al., 2019), but the applications of these early-stage methods are still limited in the biomedical field.

*1.1.4 Attention-based studies*

Another approach to bypass the ROI annotation requirement is to let the model choose the discriminative patches for survival analysis by assigning more weights to instances which are more important, which is broadly referred to as the attention mechanism. It was first designed for the sequence to sequence (seq2seq) models, such as natural machine translation, to help memorize long source sentences (Bahdanau et al., 2015). The attention mechanism was soon brought to the computer vision field to generate captions for a given image by sequentially focusing on different locations of the image (Xu et al., 2015). In the digital pathology field, a study developed by Tomita et al. used a 3-D convolutional layer to derive an attention map for aggregating feature vectors from all patches (Tomita et al., 2019). 64 such filters were used to increase the model's capacity to recognize more complex patterns. However, this method was not designed for survival prediction tasks and was unable to handle gigapixel whole-slides due to the lack of a sampling scheme.

As for the MIL-based analysis, Ilse et al. proposed an attention mechanism to adaptively aggregate multiple instances (Ilse et al., 2018). In this model, each instance is assigned an attention weight, and their respective feature vectors are combined by taking the weighted average over all instances. This attention mechanism was incorporated into the WSISA model by a later method called DeepAttnMISL (Yao et al., 2020). In this study, the patches were first grouped into multiple clusters (e.g., 8) using K-means. For each patient, the cluster-specific features were obtained from patches belonging to the same cluster. Finally, these multiple cluster-specific features were aggregated using an attention layer. This approach achieves promising results for survival prediction. However, as the patch clusters were determined beforehand, they may not be directly related to prognosis. Besides, as the clustering algorithm blocks the gradient flow from patches to the final predictions, it cannot be trained end-to-end.

A recent breakthrough in sequence analysis, the transformer model, is based solely on the attention mechanism (Vaswani et al., 2017). The transformer model consists of several layers of multi-head self-attention modules and outperforms previous methods by a large margin. The attention mechanism used in transformer is also called Query-Key-Value (QKV) attention, following the terminology in database systems. Briefly, each record is stored as a key-value pair, where the value is the actual content. The query token works as the "search" term and will be compared with the key token to determine the relevance of each record. The self-attention algorithm allows each element to "attend" to every other component in the sequence. By integrating information from each other, a better representation could be obtained. The transformer model has achieved exciting results on both natural language processing (Devlin et al., 2019) and computer vision tasks (Dosovitskiy et al., 2020). However, given the large number of parameters, it is not feasible to apply the transformer model directly to whole-slide images. This is not only due to the gigabyte size of WSIs but also because training a transformer model from scratch requires significant amounts of data and computational resources.

*1.2 Contributions*

Inspired by the powerful multi-head attention mechanism introduced in the transformer model, we innovatively modified this mechanism to allow it to work efficiently in the field of multiple instance learning. Specifically, the transformer model focuses on learning a context-sensitive representation for each element in the input sequence. This requires attention between every pair of elements and the computational cost increases quadratically with input length. However, for the prognostic information extraction task, only the global representation of the input sequence is of interest. This allows us to remove the nonessential but computationally expensive pairwise attentions among the input sequence and only keep the attentions between global query and local keys. With this modification, the computational cost scales linearly with input length.

Based on this mechanism, we present a multi-head attention framework for cancer survival prediction (MHAttnSurv) using WSIs without the need for ROI annotations. As shown in Figure 1, we first use a backbone ResNet model to extract features from randomly sampled WSI patches. Then we project the feature map into values and keys. We split the value and key matrix and the learnable query vector into several chunks. Within each chunk, the attention process works in parallel to explore and identify the discriminative regions. The result from each attention map will then be concatenated for survival prediction.

Compared to the MIL attention mechanism devised by Ilse et al., which only has one attention head (Ilse et al., 2018), our approach provides more flexibility in aggregating features from multiple patches while using the same number of parameters and maintaining the same computational cost. Moreover, compared to the DeepAttnMISL method (Yao et al., 2020), we remove the pre-processing clustering step, which makes the model easier to apply to various tasks, and enables gradient flow from the loss function to the backbone model for further finetuning. Besides, the multiple attention heads work in parallel so that the training procedure can be more efficient. Finally, we use a special dropout operation to mask out the same channels across all patients within the same batch, which is instrumental to reduce overfitting for the survival prediction task. We compare our method with current state-of-the-art methods and perform extensive experiments on 4 cancer types from The Cancer Genome Atlas (TCGA) database to demonstrate its superiority. Our contributions can be summarized as follows:

- We present an efficient and flexible attention-based framework with multiple attention heads for survival prediction.
- We performed rigorous experiments using nested cross-validation. The experiments demonstrated that our proposed method performs better than existing state-of-the-art approaches, and is easier to implement and adapt for various tasks.
- From visualizing the attention map and inspecting the prediction ability of each attention head, we have demonstrated that our approach incorporates comprehensive morphological patterns into survival prediction.

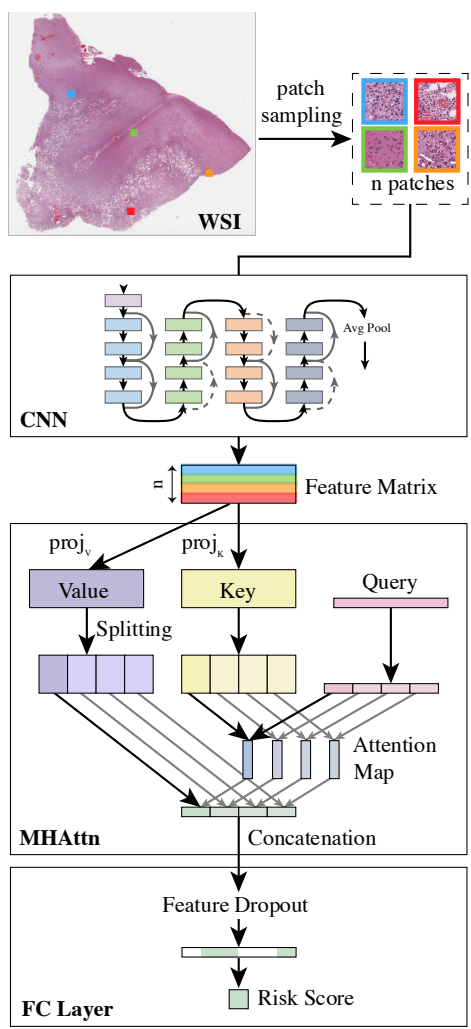

Figure 1. Overview of the model structure of MHAttnSurv. From each WSI, $n$ patches are randomly sampled. 4 heads are shown in the multi-head attention part. WSI: whole slide image; CNN: convolutional neural network; MHAttn: multi-head attention; FC layer: fully connected layer.

## 2. Methodology

For each patient $i$ in our study, the observed follow-up time is denoted by $t_i$ and the status at the end of follow-up is $\delta_i$. For a patient who died during the follow-up, $\delta_i = 1$ and $t_i$ is the survival time of this patient since cancer diagnosis. Otherwise $\delta_i = 0$ indicating the patient was censored. $P_i = \{p_1, \ldots, p_{c_i}\}$ denotes all the $c_i$ patches that belongs to this patient, and the number of patches $c_i$ differs across patients. Under the framework of MIL, the order of the patches does not matter as $P$ is considered as a bag of instances (Dietterich et al., 1997; Maron and Lozano-Pérez, 1998). Our goal is to learn a function that estimates the risk score from $P$ for every patient, so that the lower this score is, the longer this patient is expected to survive after cancer diagnosis. Our proposed MHAttnSurv method is illustrated in Figure 1.

*2.1 Sampling and feature extraction*

Because each WSI typically contains thousands of patches and each patient can have multiple WSIs, it is not feasible to feed all of them to a model at once. Instead, we use a sampling approach to select $n$ patches randomly during each iteration.

The ResNet model pre-trained on ImageNet provides good visual feature representations and is widely used in biomedical domains (Deng et al., 2009; He et al., 2015; Wei et al., 2019). We use a ResNet model with 18 layers (i.e., ResNet-18) for feature extraction from the randomly sampled patches. By using the pre-trained backbone model, we can save time and computational resources by skipping the step to develop a feature extractor and focus on developing the survival prediction model. The feature vectors from the ResNet-18 model are of size 512, and they are used as the input for the attention model.

*2.2 Multi-Head Attention*

The malignancy of a tumor is affected by many factors, including the size of the tumor and the mitotic activity of cells within the tumor. While focusing only on the tumor tissue can have a better assessment of the features of cancer cells, it will miss some global features, such as the size of the tumor. In contrast, when taking the entire slide into account, the relative size of the tumor on the slide can be reflected by the intensity of the activation signal from the neural networks, as the averaging operation will reduce the signal of a smaller tumor. However, considering the normal tissues with equal attention will dilute the impact of the tumor cells. Therefore, to reach a more accurate prediction of the patient's survival, it is necessary to inspect the histopathology slide from various aspects, and we present a novel multi-head attention model to achieve this goal.

Our proposed multi-head attention mechanism consists of three components analogous to the self-attention in a transformer model, i.e., query ($\boldsymbol{Q}$), key ($\boldsymbol{K}$), and value ($\boldsymbol{V}$). After applying the ResNet model, the feature embedding of a WSI with $n$ sampled patches is denoted as $\boldsymbol{X} \in \mathbb{R}^{n \times d}$, where $d$ is the embedding dimension of each patch (i.e., 512 as ResNet-18 is used as the backbone). $\boldsymbol{X}$ is then projected into $\boldsymbol{K}$ and $\boldsymbol{V}$. Specifically, for $\boldsymbol{V}$, we simply use an identity function, as the derived feature vectors from the pretrained ResNet model have already been adequate feature representations. The projection function for the $\boldsymbol{K}$ consists of a matrix multiplication with $\boldsymbol{W}^K \in \mathbb{R}^{d \times d}$, a dropout layer (Hinton et al., 2012), and a ReLU (Nair and Hinton, 2010) activation function. The output $\boldsymbol{K}$ can be written as,

$$\boldsymbol{K} = Relu(Dropout(\boldsymbol{X}\boldsymbol{W}^K))$$

$Q \in \mathbb{R}^{1 \times d}$ is a learnable vector with random initialization, where each element is sampled from uniform distribution $\mathcal{U}(-1/\sqrt{d}, 1/\sqrt{d})$. The attention can be written as

$$S = Attention(Q, K, V) = softmax(QK^\top)V$$

where the softmax normalization is applied along the patch dimension so that the attention weights of the $n$ patches will sum to 1. The output $S \in \mathbb{R}^{1 \times d}$ aggregates information from $n$ sampled patches and is used as the feature vector of the whole slide.

When expanding the attention layer from one to $h$ heads, we split the $Q$, $K$, and $V$ into chunks of size $d/h$ along the embedding dimension. We then calculate the attention scores within each attention head and concatenate the final representation vectors. Formally, the multi-head attention can be expressed as

$$MultiHead(Q, K, V) = concat(head_1, \ldots, head_h)$$

Where $head_i = Attention(Q_i, K_i, V_i)$ and $Q_i$ is the $i$th chunk of $Q$ and so on.

## 2.3 Feature Dropout

Dropout has been proven to be useful in preventing overfitting and improving the robustness of a model (Hinton et al., 2012). Dropout is usually applied to the hidden layers of a neural network. But given the simplicity of our multi-head attention mechanism, no hidden layer is included between the value projection layer and the final prediction layer. Instead, we apply the dropout directly on the output from the multi-head attention layer.

In the context of survival prediction, the predicted score of one patient (i.e., risk score) will be compared against all the other patients in the same batch to compute the survival loss (Section 2.4). If we randomly drop some features for each patient, it will be challenging to compare the risk scores. So, instead of using the standard dropout method, we implemented a feature-level dropout method (Figure 2). The implementation is similar to a recently developed DropHead method designed for the transformer model (Zhou et al., 2020), but the dropout is applied for randomly sampled features rather than attention heads. During each iteration, we randomly select some features for dropout, and these selected features are the same across the patients. This will ensure that we use the same subset of features for all the patients in the same batch and their risk scores are comparable.

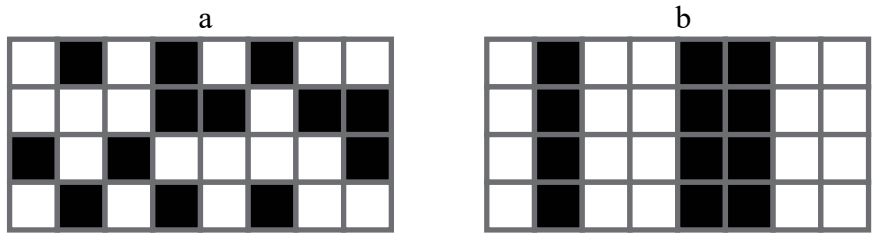

Figure 2. Illustration of the dropout method. Rows: patients; Columns: feature channels. Black cell denotes a value of 0 (i.e., masked out). (a) standard dropout method drops features for each data point independently; (b) feature dropout method drops randomly selected features for all patients.

*2.4 Loss function*

A fully connected layer was used to estimate the risk scores of the patients. We use the Cox proportional hazards function as the loss function (Cox, 1972), which is defined as

$$l(\theta) = -\frac{1}{N_{\delta=1}} \sum_{i:\delta_i=1} \left( \hat{h}_\theta(x_i) - \log \sum_{j \in \mathcal{R}(T_i)} exp\left(\widehat{h_\theta}(x_j)\right) \right)$$

where $\hat{h}_\theta(x_i)$ is the output (risk score) of a model with parameters $\theta$ and input $x_i$, $\delta = 1$ means having the event (death), and $\mathcal{R}(T_i)$ is patient $i$'s risk set (i.e., those who are alive at the time $T_i$).

# 3. Experiments and Results

*3.1 Dataset description*

The datasets that we used in our experiments are from the TCGA Program (https://www.cancer.gov/tcga). The TCGA is a public database with comprehensive clinicopathological data and multi-platform molecular profiles for more than 30 cancer types, with data collected from multiple institutions. In this study, we selected 4 cancer types, Urothelial Bladder Carcinoma (BLCA), Breast Invasive Carcinoma (BRCA), Colon Adenocarcinoma (COAD), and Lower Grade Glioma (LGG). For each cancer type, we acquired the digitized Hematoxylin and Eosin (H&E) stained slides, the follow-up information, and the survival status at the end of the follow-up. Both flash-frozen and Formalin-Fixed Paraffin-embedded (FFPE) slides were included for model development and evaluation except for LGG, for which we found the flash-frozen slides contains excessive artifacts. A detailed description of the cancer types included in this study is shown in Table 1. In total, 2,375 patients were included in the experiment from 6,162 WSIs. The largest dataset is BRCA with over 1,000 patients included, and the smallest dataset is BLCA with only 386 patients. On average, the patients were followed up by 2 years after the cancer diagnosis until they died or the last visit.

The WSIs are scanned at 20× or 40× magnification. We converted the WSIs to 10× magnification and cropped the entire slide into patches of size 224×224 without overlapping to reduce the computational and storage cost. Another benefit of using a relatively smaller magnification level is that each patch will cover a larger spatial area without losing significant cellular feature details. To determine whether a patch belongs to the background, we counted the number of purple pixels in every patch and discarded patches with less than 100 purple pixels. A total of 15.55 million foreground patches were extracted across the four cancer datasets.

|  | BLCA | BRCA | COAD | LGG |
|---|---|---|---|---|
| Number of patients | 386 | 1,050 | 449 | 490 |
| Number of WSIs | 898 | 3,003 | 1,418 | 843 |
| Number of patches (million) | 3.05 | 6.46 | 2.97 | 3.07 |
| Median follow-up time (years) | 1.65 | 2.34 | 1.80 | 1.87 |
| Number of events (deaths) | 175 | 144 | 100 | 115 |

Table 1. Descriptive statistics of the utilized datasets from TCGA.

*3.2 Implementation details*

The proposed method was implemented in Python with the PyTorch library (version 1.8.1) (Paszke et al., 2019). During each iteration, we randomly sampled 32 patches per patient for 64 patients producing 2,048 patches per batch. For a training split with 320 patients, one training epoch consists of 5 (320 patients / 64 patients per batch) parameter update steps. We evaluated the model on the validation dataset every 100 epochs (or 500 steps) to monitor the training progress, with 100 patches sampled from every patient.

Adam optimizer was used to optimize the parameters of the model (Kingma and Ba, 2015). The starting learning rate was set to $6\times10^{-5}$. A cosine learning rate scheduler was used for optimization, and we reset the learning rate every 4,000 epochs. The model was trained for up to 50,000 epochs. We monitored the c-index on the validation set to determine the early stopping epoch for each configuration. Then we evaluated the models on the test dataset.

*3.3 Evaluation metrics and methods*

The model was evaluated using the concordance index (c-index), which is widely used as an evaluation metric in time-to-event analyses. It is defined as the proportion of pairs whose rank is correctly predicted among all possible pairs (Harrell et al., 1982). C-index ranges from 0 to 1. A c-index of 1 means perfect prediction, while c-index = 0.5 means a random prediction.

As the sample size of the chosen cancer types is only several hundred, if we allocate 70% of the cases for model training and validation and the remaining 30% for model testing (as this is typically practiced in the deep learning field), we will end up with too few death events in our test set, and the c-index could be highly stochastic and unreliable. To overcome this problem, we used the nested cross-validation (Nested-CV) method to obtain a more stable estimate of the model's performance. The Nested-CV method makes the most use of the data while avoiding information leakage and is widely used in evaluating machine learning methods with a moderate sample size (Raschka, 2018). Briefly, our Nested-CV framework consists of 5 outer loops, and each outer loop consists of 4 inner loops. The inner loop is essentially $K$-fold cross-validation for hyper-parameter tuning. The outer loop is used to obtain the model's performance estimate on all patients, which is particularly beneficial to obtain a more stable estimate of the performance of the proposed method when only a limited number of events/deaths are available.

As the death events are only observed for part of the study participants, to avoid possible imbalance, we split the datasets by stratifying on the outcome status of the patient, i.e., whether the patient was alive or not at the last follow-up. Both data splitting and model evaluation was performed at the patient level to avoid any potential information leak. We finetuned one hyper-parameter, the dropout rate for the final layer. We used a grid search approach to evaluate several dropout rates, namely, {0.0, 0.2, 0.5, 0.8, 0.95}. For each outer loop, we selected the dropout rate with the best average validation c-index.

1,000 patches were randomly sampled for each patient in the test splits to evaluate the selected model. For each outer loop, we averaged the predictions from each inner loop and calculated the c-index. For the overall performance, we reported the average c-index across multiple outer loops.

*3.4 Quantitative comparison with other methods*

Based on our ablation study (Section 3.6), we set the number of attention heads to 8. We compared our model with two other state-of-the-art methods, the average pooling method (AvgPool) (Wulczyn et al., 2020) and DeepAttnMISL (Yao et al., 2020). For the DeepAttnMISL method, we used the same number of clusters as the number of attention heads (i.e., 8). In the original DeepAttnMISL study, the performance of the model with 8 clusters is very close to that of the best model (either 6 or 10 clusters). We used the same strategy for hyperparameter tuning and model evaluation for the baseline methods as described in Section 3.3. And we fixed the same split when repeating the nested cross-validation for different models. Because the outer loop specific c-index can be viewed as a repeated measurement of the model performance across model types, we fit linear mixed models in R using the lme4 package (Bates et al., 2014) to compare the performance of different models. Specifically, the c-index was used as the outcome, and the model's name was used as the predictor. Cancer type and outer fold as modeled as random effects with outer fold nested within cancer type.

The comparison results are shown in Table 2. Our model achieved the best results for all four cancer types selected. Specifically, the differences between our method and the AvgPool method were statistically significant for the BLCA dataset ($p < 0.01$) and marginally significant ($p < 0.1$) for the LGG dataset. Overall, the average c-index of our model is 0.640, which is significantly better than the AvgPool method (c-index = 0.603, $\Delta = 0.037$). While our method achieved a consistently higher c-index compared to the DeepAttnMISL method, the difference was not statistically significant. The average c-index of the DeepAttnMISL method was 0.619, which was lower than the proposed MHAttnSurv method by 0.021 ($p = 0.05$).

We further examined the time-varying AUC to assess the performance of each model up to 5 years after diagnosis. At each assessed time point, this evaluation method considered the patients who died before this time point as cases, and patients who were still alive after this time point as controls, to measure the discrimination power of the model predictions. Inverse probability weighting is used to correct for right censoring (Blanche et al., 2013). Figure 3 shows that the models are typically better at predicting early events. For all cancers except LGG, our MHAttnSurv method performs the best for almost all the evaluated time points. While for LGG, our method performs better than both AvgPool and DeepAttnMISL up to 3 years after diagnosis. For the last two time points, the AUC is lower than DeepAttnMISL by about 0.02. Overall, the MHAttnSurv method is the best of the three in terms of time-varying AUC.

Additionally, we included the Kaplan-Meier curves for each model in Supplementary Figure S1. For the test dataset of each outer fold, we split the predicted risk score into tertiles (i.e., based on the 33$^{rd}$ and 67$^{th}$ percentile) and denote them as the low-risk group, medium-risk group, and high-risk group. The log-rank test p-values of the MHAttnSurv method are significant for all four cancer types, and there is a visible separation in Kaplan-Meier curves between the high-risk group with the low/medium group.

|  | BLCA | BRCA | COAD | LGG | ALL |
| --- | --- | --- | --- | --- | --- |
| AvgPool | 0.551 | 0.575 | 0.601 | 0.685 | 0.603 |
| DeepAttnMISL | 0.594 | 0.600 | 0.581 | 0.700 | 0.619 |
| MHAttnSurv (ours) | **0.604** | **0.607** | **0.636** | **0.714** | **0.640** |
| Δ1: MHAttnSurv – AvgPool | 0.054** | 0.031 | 0.035 | 0.029† | 0.037** |
| Δ2: MHAttnSurv – DeepAttnMISL | 0.010 | 0.007 | 0.055 | 0.013 | 0.021† |

Table 2. Comparison of model performance. **Bold**: Best c-index. †: p-value < 0.1; *: p-value < 0.05; **: p-value < 0.01.

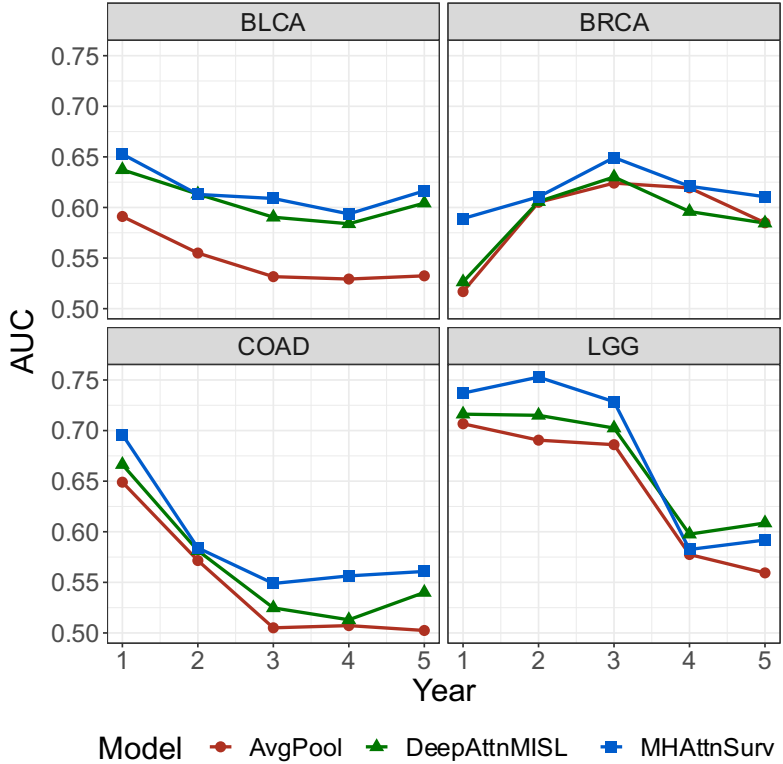

Figure 3. Time-varying AUC evaluated annually within 5 years after diagnosis.

*3.5 Qualitative and Quantitative Analysis of MHAttnSurv*

To better understand how the attention model works, we present the connections between attention heads and patch features for one example WSI from the LGG dataset in Figure 4. Specifically, Figure 4(a) illustrates the attention map for each individual attention head, and Figure 4(b) shows the cluster membership of each patch as well as two example patches of each cluster. To derive the head-wise attention maps, we first calculated the patch-level attention score for all the patches across 8 attention heads. Then iteratively, we randomly sampled 32 patches and calculated the attention weights from attention scores using the softmax function. We repeated this process so every patch was sampled 10 times and we calculated the average attention weights for each patch. The attention weights are rescaled by multiplying by 32.

As shown in Figure 4, the attention maps vary widely across attention heads. For example, Head-5 (H5) focuses mainly on regions with patches from Cluster-2 (C2), C6, and C7, characterized by high nucleus density and active angiogenesis. H7 focuses on a similar region as H5 but with stronger attention. On the contrary, H1 concentrates mainly on patches from C1, which are most likely from normal tissues. Interestingly, H8 focuses primarily on the edges where the tissues coexist with the background. The rich attention patterns enable the MHAttnSurv model to comprehensively consider various morphological features which might contribute to prognosis. Additional example predictions for the other cancer types are shown in Supplementary Figure S2, S3, and S4.

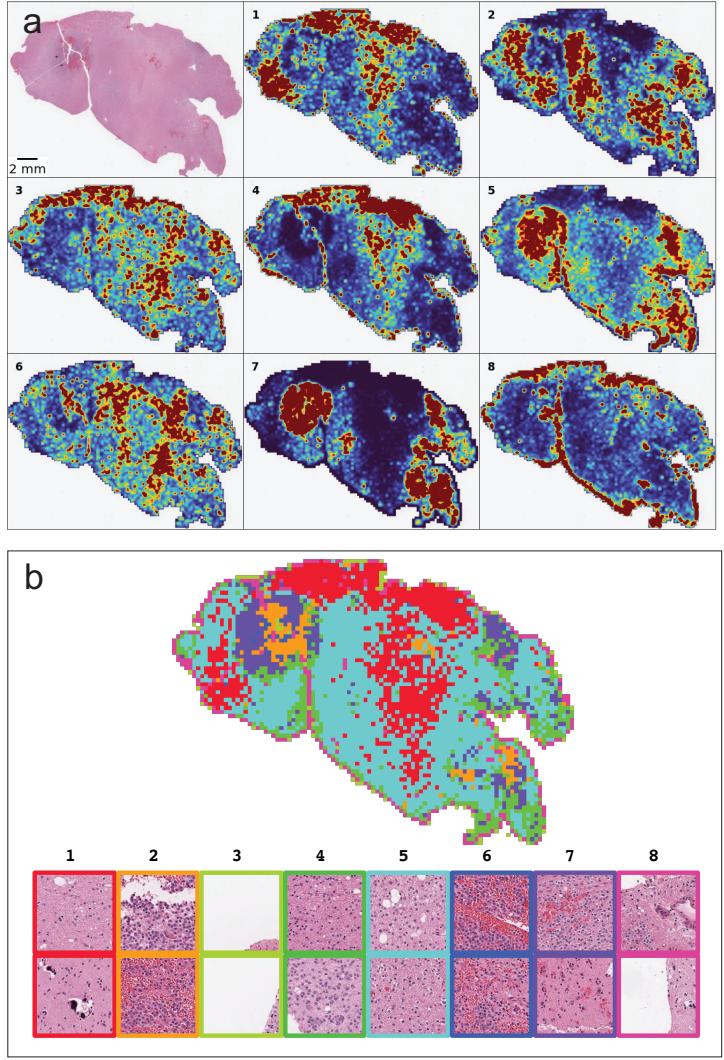

Figure 4. Visualization of head-wise attention map and patch clusters for one sample WSI from LGG. (a) Head-wise attention map. Red color: rescaled-attention weights > 2; Blue color: rescaled-attention weights = 0. (b) Patch clusters on the WSI level and example patches from each cluster.

To illustrate the contribution of each attention head to the overall prediction, we calculated the c-index for each attention head. For example, when inspecting H1, we zeroed out all the other heads and only kept H1, and calculated the c-index based on the predictions from this head. We present the head-wise results in Table 3, along with the results when keeping all attention heads. These results indicate that the performance varies across attention heads, with the best head-wise c-index observed for H1, H4, H3, and H2, for BLCA, BRCA, COAD, and LGG, respectively. Of note, using all the attention heads in our approach outperformed using each single attention head across all four cancer types. The results suggest that multiple attention heads provide complementary information to achieve a better representation of the prognostic features.

| Head | BLCA | BRCA | COAD | LGG |
|---|---|---|---|---|
| 1 | <u>0.551</u> | 0.511 | 0.512 | 0.585 |
| 2 | 0.548 | 0.561 | 0.503 | <u>0.690</u> |
| 3 | 0.511 | 0.545 | <u>0.615</u> | 0.630 |
| 4 | 0.510 | <u>0.601</u> | 0.545 | 0.650 |
| 5 | 0.521 | 0.580 | 0.573 | 0.582 |
| 6 | 0.495 | 0.507 | 0.554 | 0.663 |
| 7 | 0.507 | 0.571 | 0.590 | 0.544 |
| 8 | 0.542 | 0.560 | 0.533 | 0.574 |
| All | **0.604** | **0.607** | **0.636** | **0.714** |

Table 3. Head-wise c-index. The "All" row refers to keeping all heads predictions. Best results for each column are denoted with **boldface**. The second-best results are <u>underscored</u>.

We further explored the correlation among the attention heads for LGG as showed in Figure 5. We can see that the attention weights across different heads are weakly correlated, further demonstrating that each attention head focuses on distinct patterns from a whole-slide image. On the contrary, there exist much stronger correlations among patch-level risk scores across attention heads. This is interesting as the head-wise patch-level risk scores are each calculated using a distinct chunk of the extracted feature vector from the ResNet-18 model. Meanwhile, we noticed that the head-wise c-index could be very different for two attention heads with a strong correlation in patch-level risk scores. Taking H2 and H7 for example, the correlation coefficients of their patch-level attention weights and patch-level risk scores are 0.24 and 0.67, respectively. Despite the high correlation coefficient in their patch-level risk score predictions, their head-wise c-index are very different (0.690 vs. 0.544). This suggests the attention pattern is a driving force in the performance of head-wise predictions.

a

| | $head_1$ | $head_2$ | $head_3$ | $head_4$ | $head_5$ | $head_6$ | $head_7$ | $head_8$ |
|---|---|---|---|---|---|---|---|---|
| $head_1$ | 1.00 | -0.09 | 0.09 | 0.15 | -0.06 | -0.09 | -0.12 | -0.02 |
| $head_2$ | -0.09 | 1.00 | 0.05 | 0.00 | 0.12 | 0.41 | 0.24 | -0.08 |
| $head_3$ | 0.09 | 0.05 | 1.00 | 0.33 | -0.09 | 0.04 | -0.12 | 0.19 |
| $head_4$ | 0.15 | 0.00 | 0.33 | 1.00 | -0.02 | 0.06 | -0.13 | 0.12 |
| $head_5$ | -0.06 | 0.12 | -0.09 | -0.02 | 1.00 | 0.19 | 0.37 | 0.08 |
| $head_6$ | -0.09 | 0.41 | 0.04 | 0.06 | 0.19 | 1.00 | 0.13 | 0.01 |
| $head_7$ | -0.12 | 0.24 | -0.12 | -0.13 | 0.37 | 0.13 | 1.00 | -0.07 |
| $head_8$ | -0.02 | -0.08 | 0.19 | 0.12 | 0.08 | 0.01 | -0.07 | 1.00 |

b

| | $head_1$ | $head_2$ | $head_3$ | $head_4$ | $head_5$ | $head_6$ | $head_7$ | $head_8$ |
|---|---|---|---|---|---|---|---|---|
| $head_1$ | 1.00 | -0.88 | 0.55 | -0.53 | -0.68 | -0.84 | -0.70 | 0.42 |
| $head_2$ | -0.88 | 1.00 | -0.50 | 0.45 | 0.66 | 0.78 | 0.67 | -0.35 |
| $head_3$ | 0.55 | -0.50 | 1.00 | 0.15 | -0.52 | -0.45 | -0.53 | 0.49 |
| $head_4$ | -0.53 | 0.45 | 0.15 | 1.00 | 0.14 | 0.51 | 0.17 | -0.09 |
| $head_5$ | -0.68 | 0.66 | -0.52 | 0.14 | 1.00 | 0.59 | 0.78 | -0.27 |
| $head_6$ | -0.84 | 0.78 | -0.45 | 0.51 | 0.59 | 1.00 | 0.59 | -0.22 |
| $head_7$ | -0.70 | 0.67 | -0.53 | 0.17 | 0.78 | 0.59 | 1.00 | -0.19 |
| $head_8$ | 0.42 | -0.35 | 0.49 | -0.09 | -0.27 | -0.22 | -0.19 | 1.00 |

Figure 5. Correlation of attention weights and risk scores across attention heads for LGG. (a) correlation of attentions weights. (b) correlation of patch risk scores.

*3.6 Effects of number of attention heads*

| Number of heads | BLCA | BRCA | COAD | LGG | ALL |
|---|---|---|---|---|---|
| 1 | 0.576 | 0.592 | 0.583 | 0.685 | 0.609 |
| 4 | 0.577 | **0.614** | 0.628 | 0.702 | 0.630 |
| 8 | **0.604** | 0.607 | **0.636** | 0.714 | **0.640** |
| 16 | 0.567 | 0.614 | 0.626 | 0.710 | 0.629 |
| 32 | 0.583 | 0.570 | 0.626 | **0.715** | 0.623 |

Table 4. Effect of the number of attention heads.

As the ablation study, we explored whether using a different number of attention heads could affect the model performance. The results are presented in Table 4. When using only one attention head, the average c-index is 0.609, which is comparable to (or slightly better than) the AvgPool method (c-index: 0.603). However, with 4 attention heads, the average c-index increased to 0.630. Using 8 attention heads achieves the best overall results, but using more attention heads (e.g., 16 and 32) results in performance degradation. For BLCA and COAD datasets, the best results are observed when using 8 attention heads. For BRCA and LGG, the best results occur when using 4 and 32 attention heads, respectively. Based on our results in this ablation study, starting with 8 attention heads is a reasonable choice in survival prediction studies.

# 4. Discussion

In this work, we presented a novel multi-head attention approach for survival prediction on whole-slide pathology images. Accurate survival prediction and cancer risk stratification can benefit the cancer patients for making decisions on treatment plans and better physical and psychological adaptation (Hagerty et al., 2005; Weeks et al., 1998). An automated and effective prognosis prediction program will also benefit the physicians and pathologists by reducing their workload and allowing them to communicate more confidently with the patients when discussing treatment options (Christakis and Iwashyna, 1998).

Prognosis prediction has been a challenging problem. Unlike the tumor subtype classification and tumor segmentation tasks, where the histopathological aspects are well-defined, making accurate prognosis prediction based on histology slides is difficult. A comprehensive prognosis evaluation requires careful consideration of both the global and local structures of the pathology slide. This problem is exacerbated by the large size of WSIs. Most of the deep learning methods developed so far either require ROI annotations (Mobadersany et al., 2018; Zhu et al., 2017a) or make strong assumptions that every patch contributes equally to the prognosis prediction (Wulczyn et al., 2020). ROI annotations are costly to obtain, and the annotation process is subjective to pathologists' discretion. While for approaches that assign equal weights to each patch, the cancer malignancy signal carried by the tumor tissue can be easily diluted by surrounding normal tissues.

The experimental results demonstrated the effectiveness of our proposed multi-head attention mechanism, in comparison to two strong deep learning baseline approaches, AvgPool and DeepAttnMISL. The improvement in the performance of our model can be explained by the following reasons. Firstly, our model employs the attention mechanism to combine patches based on their relevance to the overall task. This strategy could avoid assigning the same weight (as in the AvgPool method) to patches that are not prognostically relevant. Secondly, the multiple attention heads can achieve the same effect as the siamese model used in the DeepAttnMISL. Each attention head can provide some unique aspects regarding how to combine the different patches. But unlike the fixed clustering algorithm used in DeepAttnMISL, the attention mechanism used in our model is more flexible. The different attention heads are not determined ahead of time from the extracted patch features. Instead, each head is allowed to focus freely on patterns that collectively benefit most to the prognosis prediction. Finally, we introduced a feature dropout method in our approach, which is specifically designed for the survival prediction task and can improve the generalizability of the model in the absence of large datasets.

The visualization of the attention maps demonstrates that each attention head acts relatively independently on exploring the regions that would benefit the overall prognosis prediction (Figure 4). While linking the attention maps to the head-wise c-index (Table 3), we note even an attention head which focuses primarily on the normal tissues (e.g., H1) can still provide some prognostic merits. One possible explanation is that the presence of normal tissue on a WSI is a positive sign for better survival. This also highlights the importance of a holistic view of the WSI in order to achieve more accurate prognosis prediction accuracy. In this study, we observe a 0.037 improvement in c-index comparing our method to the AvgPool approach. On the contrary, by focusing only on the patches within the ROI region, the c-index was improved less than 0.010 in a previous study (Wulczyn et al., 2021). This implies that even with expensive ROI annotation, our results will likely be worse without using the MHAttnSurv approach. In fact, when only one attention head is used, the average c-index of our method is only slightly better than the AvgPool approach. This observation suggests that the monochromatic ROI annotation is less effective than the multi-head attention approach for a complex task such as prognosis prediction, and further justifies our design of the multi-head attention mechanism.

As for limitations, we only evaluated the multi-head attention mechanism on one outcome type, which is survival after cancer diagnosis. The utility of this method for other outcome types is not explored. In addition, we experimented whether finetuning the backbone model can further improve the model performance. But we did not observe any noticeable improvement on the validation c-index when finetuning the backbone model. Based on our observations, the model quickly overfitted in our experiments. Overfitting persists even when only finetuning the last few convolutional layers of ResNet. This shows that although our model is trainable end-to-end, it requires a much larger dataset for effective finetuning of the backbone model.

In future work, we plan to evaluate our MHAttnSurv method with other outcome types for patients, such as binary outcomes using cross-entropy loss (e.g., mutation status of certain genes) or continuous outcomes using mean squared error loss (e.g., expression level of certain genes). We also plan to evaluate the performance of our model by training it end-to-end on larger datasets.

# 5. Conclusion

In this paper, we proposed a multi-head attention mechanism to extract prognostic information from whole-slide images that does not require the resource-intensive region of interest annotations. Compared to two existing state-of-the-art methods, our model achieved better performance on 4 TCGA datasets in terms of both c-index and time-varying AUC. Moreover, our method does not require the clustering step and is easy to implement. Visualization of the attention maps learned by our method demonstrated that each attention head focuses differently on the same whole-slide while together they work synergistically to achieve a comprehensive representation for prognosis prediction. We expect future studies to validate our method with larger datasets to further demonstrate its potentials.

# CRediT Authorship Contribution Statement

**Shuai Jiang:** Conceptualization, Methodology, Software, Investigation, Data Curation, Writing – Original Draft, Writing – Review & Editing, Visualization; **Arief A. Suriawinata:** Writing – Review & Editing, Visualization; **Saeed Hassanpour:** Conceptualization, Methodology, Writing – Review & Editing, Supervision, Project administration, Funding acquisition.

# Acknowledgments

We thank Naofumi Tomita and Joseph DiPalma for providing revision suggestions to improve the manuscript. We thank Lamar D. Moss for proofreading the article.

# Funding

This research was supported in part by grants from the US National Library of Medicine (R01LM012837) and the US National Cancer Institute (R01CA249758).

# Declaration of Competing Interests

No conflict of interest.

# Reference

American Cancer Society, 2021. American Cancer Society: Cancer Facts and Figures 2021. Atlanta Am. Cancer Soc.

Bahdanau, D., Cho, K.H., Bengio, Y., 2015. Neural machine translation by jointly learning to align and translate, in: 3rd International Conference on Learning Representations, ICLR 2015 - Conference Track Proceedings.

Bates, D., Mächler, M., Bolker, B., Walker, S., 2014. Fitting Linear Mixed-Effects Models using lme4. eprint arXiv:1406.5823 67, 51. https://doi.org/10.18637/jss.v067.i01

Betge, J., Pollheimer, M.J., Lindtner, R.A., Kornprat, P., Schlemmer, A., Rehak, P., Vieth, M., Hoefler, G., Langner, C., 2012. Intramural and extramural vascular invasion in colorectal cancer: prognostic significance and quality of pathology reporting. Cancer 118, 628–638. https://doi.org/10.1002/cncr.26310

Blanche, P., Dartigues, J.F., Jacqmin-Gadda, H., 2013. Estimating and comparing time-dependent areas under receiver operating characteristic curves for censored event times with competing risks. Stat. Med. 32, 5381–5397. https://doi.org/10.1002/sim.5958

Caicedo, J.C., Cooper, S., Heigwer, F., Warchal, S., Qiu, P., Molnar, C., Vasilevich, A.S., Barry, J.D., Bansal, H.S., Kraus, O., Wawer, M., Paavolainen, L., Herrmann, M.D., Rohban, M., Hung, J., Hennig, H., Concannon, J., Smith, I., Clemons, P.A., Singh, S., Rees, P., Horvath, P., Linington, R.G., Carpenter, A.E., 2017. Data-analysis strategies for image-based cell profiling. Nat. Methods 14, 849–863. https://doi.org/10.1038/nmeth.4397

Carpenter, A.E., Jones, T.R., Lamprecht, M.R., Clarke, C., Kang, I.H., Friman, O., Guertin, D.A., Chang, J.H., Lindquist, R.A., Moffat, J., Golland, P., Sabatini, D.M., 2006. CellProfiler: image analysis software for identifying and quantifying cell phenotypes. Genome Biol. 7, R100. https://doi.org/10.1186/gb-2006-7-10-r100

Cheerla, A., Gevaert, O., 2019. Deep learning with multimodal representation for pancancer prognosis prediction, in: Bioinformatics. pp. i446–i454. https://doi.org/10.1093/bioinformatics/btz342

Chen, S., Zhang, N., Jiang, L., Gao, F., Shao, J., Wang, T., Zhang, E., Yu, H., Wang, X., Zheng, J., 2021. Clinical use of a machine learning histopathological image signature in diagnosis and survival prediction of clear cell renal cell carcinoma. Int. J. cancer 148, 780–790. https://doi.org/10.1002/ijc.33288

Christakis, N.A., Iwashyna, T.J., 1998. Attitude and self-reported practice regarding prognostication in a national sample of internists. Arch. Intern. Med. 158, 2389–2395. https://doi.org/10.1001/archinte.158.21.2389

Cox, D.R., 1972. Regression Models and Life-Tables. J. R. Stat. Soc. Ser. B 34, 187–220.

Deng, J., Dong, W., Socher, R., Li, L., Li, K., Fei-Fei, L., 2009. ImageNet: A large-scale hierarchical image database, in: 2009 IEEE Conference on Computer Vision and Pattern Recognition. pp. 248–255. https://doi.org/10.1109/CVPR.2009.5206848

Devlin, J., Chang, M.W., Lee, K., Toutanova, K., 2019. BERT: Pre-training of deep bidirectional transformers for language understanding. NAACL HLT 2019 - 2019 Conf. North Am. Chapter Assoc. Comput. Linguist. Hum. Lang. Technol. - Proc. Conf. 1, 4171–4186.

Dietterich, T.G., Lathrop, R.H., Lozano-Pérez, T., 1997. Solving the multiple instance problem with axis-parallel rectangles. Artif. Intell. 89, 31–71. https://doi.org/https://doi.org/10.1016/S0004-3702(96)00034-3

Dosovitskiy, A., Beyer, L., Kolesnikov, A., Weissenborn, D., Zhai, X., Unterthiner, T.,

Dehghani, M., Minderer, M., Heigold, G., Gelly, S., Uszkoreit, J., Houlsby, N., 2020. An Image is Worth 16x16 Words: Transformers for Image Recognition at Scale.

Feng, Q., May, M.T., Ingle, S., Lu, M., Yang, Z., Tang, J., 2019. Prognostic Models for Predicting Overall Survival in Patients with Primary Gastric Cancer: A Systematic Review. Biomed Res. Int. 2019, 5634598. https://doi.org/10.1155/2019/5634598

Gospodarowicz, M., O'Sullivan, B., 2003. Prognostic factors in cancer. Semin. Surg. Oncol. 21, 13–18. https://doi.org/10.1002/ssu.10016

Hagerty, R.G., Butow, P.N., Ellis, P.M., Dimitry, S., Tattersall, M.H.N., 2005. Communicating prognosis in cancer care: a systematic review of the literature. Ann. Oncol. Off. J. Eur. Soc. Med. Oncol. 16, 1005–1053. https://doi.org/10.1093/annonc/mdi211

Harrell, F.E.J., Califf, R.M., Pryor, D.B., Lee, K.L., Rosati, R.A., 1982. Evaluating the yield of medical tests. JAMA 247, 2543–2546.

He, K., Zhang, X., Ren, S., Sun, J., 2015. Deep Residual Learning for Image Recognition. arXiv.

Hinton, G.E., Srivastava, N., Krizhevsky, A., Sutskever, I., Salakhutdinov, R.R., 2012. Improving neural networks by preventing co-adaptation of feature detectors 1–18.

Ilse, M., Tomczak, J.M., Welling, M., 2018. Attention-based deep multiple instance learning, in: 35th International Conference on Machine Learning, ICML 2018. pp. 3376–3391.

Jiang, S., Zanazzi, G.J., Hassanpour, S., 2021. Predicting prognosis and IDH mutation status for patients with lower-grade gliomas using whole slide images. Sci. Rep. 11, 16849. https://doi.org/10.1038/s41598-021-95948-x

Kingma, D.P., Ba, J.L., 2015. Adam: A method for stochastic optimization, in: 3rd International Conference on Learning Representations, ICLR 2015 - Conference Track Proceedings.

Li, R., Yao, J., Zhu, X., Li, Y., Huang, J., 2018. Graph CNN for Survival Analysis on Whole Slide Pathological Images. https://doi.org/10.1007/978-3-030-00934-2_20

Liu, H., Zhou, H., Yan, L., Ye, T., Lu, H., Sun, X., Ye, Z., Xu, H., 2017. Prognostic significance of six clinicopathological features for biochemical recurrence after radical prostatectomy: a systematic review and meta-analysis. Oncotarget 9, 32238–32249. https://doi.org/10.18632/oncotarget.22459

Maron, O., Lozano-Pérez, T., 1998. A Framework for Multiple-Instance Learning, in: Jordan, M., Kearns, M., Solla, S. (Eds.), Advances in Neural Information Processing Systems. MIT Press.

Mobadersany, P., Yousefi, S., Amgad, M., Gutman, D.A., Barnholtz-Sloan, J.S., Velázquez Vega, J.E., Brat, D.J., Cooper, L.A.D., Designed, L.A.D.C., Performed, L.A.D.C., 2018. Predicting cancer outcomes from histology and genomics using convolutional networks. https://doi.org/10.1073/pnas.1717139115

Morris, E.J.A., Maughan, N.J., Forman, D., Quirke, P., 2007. Who to treat with adjuvant therapy in Dukes B/stage II colorectal cancer? The need for high quality pathology. Gut 56, 1419–1425. https://doi.org/10.1136/gut.2006.116830

Nair, V., Hinton, G.E., 2010. Rectified linear units improve Restricted Boltzmann machines, in: ICML 2010 - Proceedings, 27th International Conference on Machine Learning. pp. 807–814.

Paszke, A., Gross, S., Massa, F., Lerer, A., Bradbury Google, J., Chanan, G., Killeen, T., Lin, Z., Gimelshein, N., Antiga, L., Desmaison, A., Xamla, A.K., Yang, E., Devito, Z., Raison Nabla, M., Tejani, A., Chilamkurthy, S., Ai, Q., Steiner, B., Facebook, L.F., Facebook, J.B., Chintala, S., 2019. PyTorch: An Imperative Style, High-Performance Deep Learning Library.

Phung, M.T., Tin Tin, S., Elwood, J.M., 2019. Prognostic models for breast cancer: a systematic review. BMC Cancer 19, 230. https://doi.org/10.1186/s12885-019-5442-6

Raschka, S., 2018. Model evaluation, model selection, and algorithm selection in machine learning. arXiv.

Rosenberg, J., Chia, Y.L., Plevritis, S., 2005. The effect of age, race, tumor size, tumor grade, and disease stage on invasive ductal breast cancer survival in the U.S. SEER database. Breast Cancer Res. Treat. 89, 47–54. https://doi.org/10.1007/s10549-004-1470-1

Saillard, C., Schmauch, B., Laifa, O., Moarii, M., Toldo, S., Zaslavskiy, M., Pronier, E., Laurent, A., Amaddeo, G., Regnault, H., Sommacale, D., Ziol, M., Pawlotsky, J., Mulé, S., Luciani, A., Wainrib, G., Clozel, T., Courtiol, P., Calderaro, J., 2020. Predicting survival after hepatocellular carcinoma resection using deep-learning on histological slides. Hepatology 0–3. https://doi.org/10.1002/hep.31207

Tang, B., Li, A., Li, B., Wang, M., 2019. CapSurv: Capsule Network for Survival Analysis With Whole Slide Pathological Images. IEEE Access 7, 26022–26030. https://doi.org/10.1109/ACCESS.2019.2901049

Tomita, N., Abdollahi, B., Wei, J., Ren, B., Suriawinata, A., Hassanpour, S., 2019. Attention-Based Deep Neural Networks for Detection of Cancerous and Precancerous Esophagus Tissue on Histopathological Slides. JAMA Netw. Open 2, e1914645. https://doi.org/10.1001/jamanetworkopen.2019.14645

Van Diest, P.J., Van Der Wall, E., 2004. Prognostic value of proliferation in invasive breast cancer: a review. J Clin Pathol 57, 675–681. https://doi.org/10.1136/jcp.2003.010777

Vaswani, A., Shazeer, N., Parmar, N., Uszkoreit, J., Jones, L., Gomez, A.N., Kaiser, Ł., Polosukhin, I., 2017. Attention is all you need. Adv. Neural Inf. Process. Syst. 2017-Decem, 5999–6009.

Weeks, J.C., Cook, E.F., O'Day, S.J., Peterson, L.M., Wenger, N., Reding, D., Harrell, F.E., Kussin, P., Dawson, N. V., Connors, A.F., Lynn, J., Phillips, R.S., 1998. Relationship between cancer patients' predictions of prognosis and their treatment preferences. J. Am. Med. Assoc. 279, 1709–1714. https://doi.org/10.1001/jama.279.21.1709

Wei, J.W., Tafe, L.J., Linnik, Y.A., Vaickus, L.J., Tomita, N., Hassanpour, S., 2019. Pathologist-level classification of histologic patterns on resected lung adenocarcinoma slides with deep neural networks. Sci. Rep. 9, 1–8. https://doi.org/10.1038/s41598-019-40041-7

Wulczyn, E., Steiner, D.F., Moran, M., Plass, M., Reihs, R., Tan, F., Flament-Auvigne, I., Brown, T., Regitnig, P., Chen, P.-H.C., Hegde, N., Sadhwani, A., MacDonald, R., Ayalew, B., Corrado, G.S., Peng, L.H., Tse, D., Müller, H., Xu, Z., Liu, Y., Stumpe, M.C., Zatloukal, K., Mermel, C.H., 2021. Interpretable survival prediction for colorectal cancer using deep learning. npj Digit. Med. 4, 71. https://doi.org/10.1038/s41746-021-00427-2

Wulczyn, E., Steiner, D.F., Xu, Z., Sadhwani, A., Wang, H., Flament-Auvigne, I., Mermel, C.H., Chen, P.H.C., Liu, Y., Stumpe, M.C., 2020. Deep learning-based survival prediction for multiple cancer types using histopathology images. PLoS One 15, 1–18. https://doi.org/10.1371/journal.pone.0233678

Xu, K., Ba, J.L., Kiros, R., Cho, K., Courville, A., Salakhutdinov, R., Zemel, R.S., Bengio, Y., 2015. Show, attend and tell: Neural image caption generation with visual attention, in: 32nd International Conference on Machine Learning, ICML 2015. pp. 2048–2057.

Yao, J., Zhu, X., Huang, J., 2019. Deep Multi-instance Learning for Survival Prediction from Whole Slide Images. https://doi.org/10.1007/978-3-030-32239-7_55

Yao, J., Zhu, X., Jonnagaddala, J., Hawkins, N., Huang, J., 2020. Whole slide images based

cancer survival prediction using attention guided deep multiple instance learning networks. Med. Image Anal. 65, 101789. https://doi.org/10.1016/j.media.2020.101789

Zhou, W., Ge, T., Wei, F., Zhou, M., Xu, K., 2020. Scheduled DropHead: A Regularization Method for Transformer Models. pp. 1971–1980. https://doi.org/10.18653/v1/2020.findings-emnlp.178

Zhu, X., Yao, J., Huang, J., 2017a. Deep convolutional neural network for survival analysis with pathological images. Proc. - 2016 IEEE Int. Conf. Bioinforma. Biomed. BIBM 2016 544–547. https://doi.org/10.1109/BIBM.2016.7822579

Zhu, X., Yao, J., Zhu, F., Huang, J., 2017b. WSISA: Making survival prediction from whole slide histopathological images. Proc. - 30th IEEE Conf. Comput. Vis. Pattern Recognition, CVPR 2017 2017-Janua, 6855–6863. https://doi.org/10.1109/CVPR.2017.725

**Supplementary Materials**

Figure S1. Kaplan-Meier curves comparing MHAttnSurv method with AvgPool and DeepAttnMISL methods.

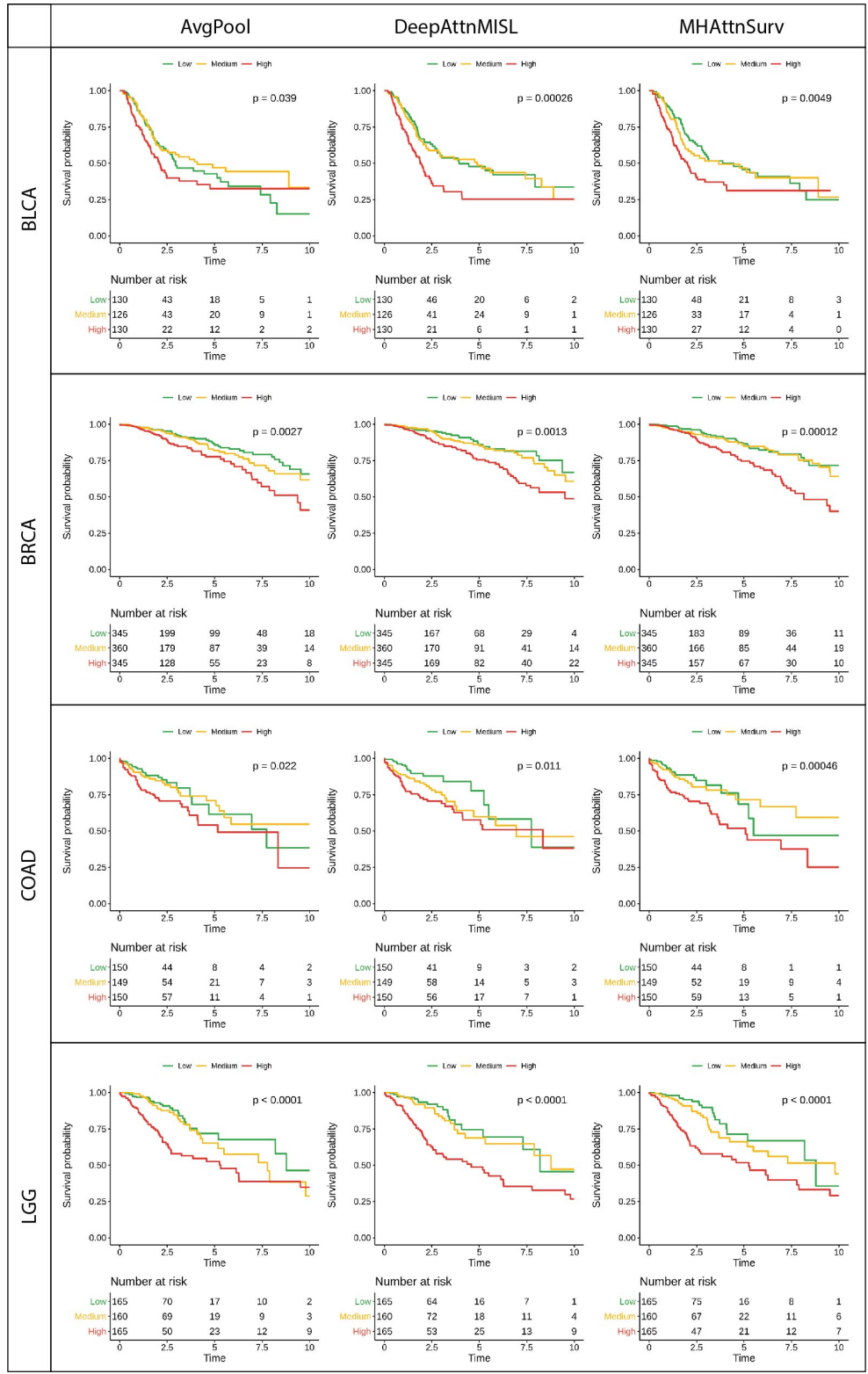

Supplementary Figure S2. Visualization of head-wise attention map and patch clusters for one sample WSI from BLCA. (a) Head-wise attention map. Red color: rescaled-attention weights > 2; Blue color: rescaled-attention weights = 0. (b) Patch clusters on the WSI level and example patches from each cluster. No patch belongs to cluster 6 for this chosen WSI.

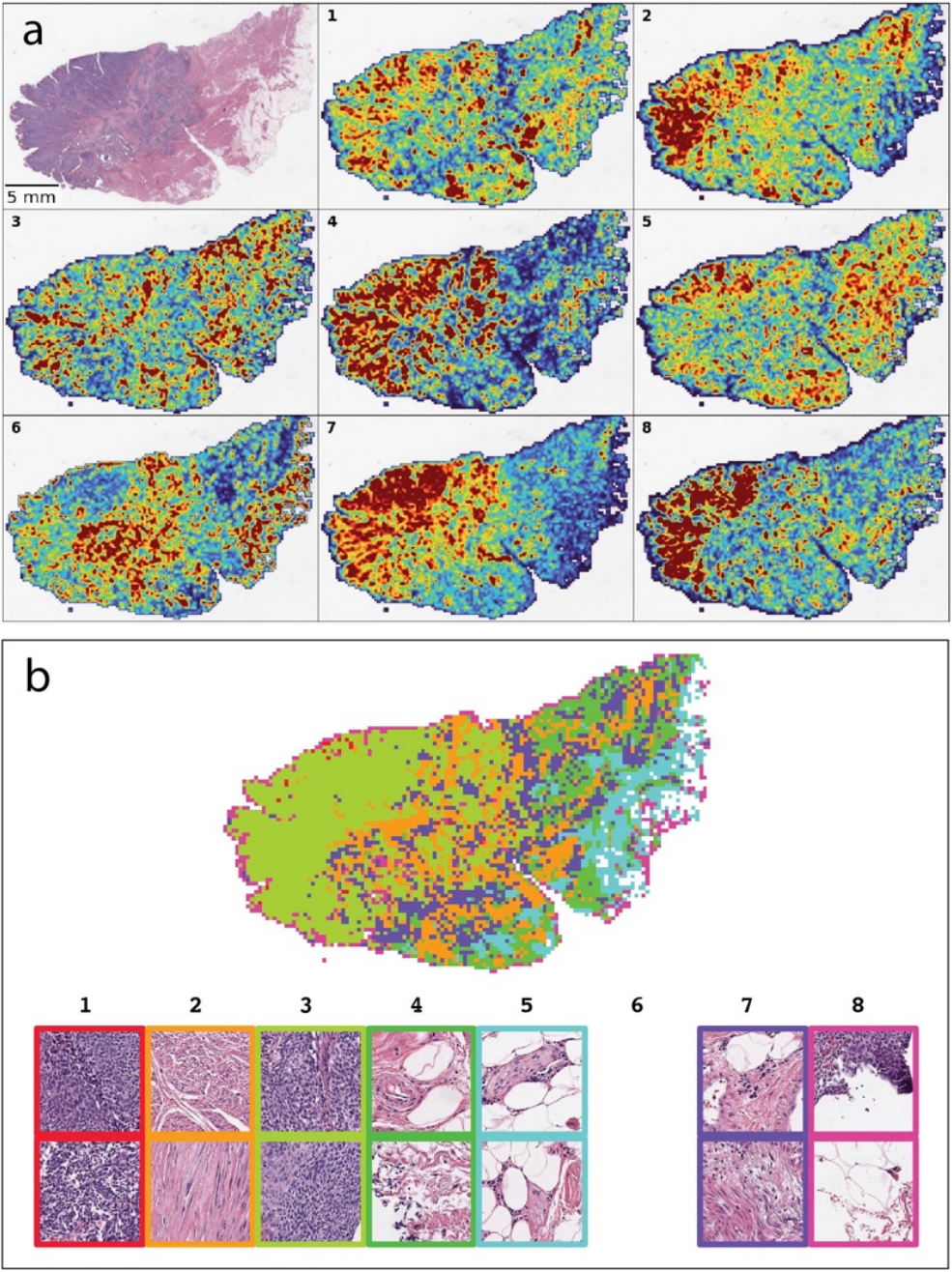

Supplementary Figure S3. Visualization of head-wise attention map and patch clusters for one sample WSI from BRCA. (a) Head-wise attention map. Red color: rescaled-attention weights > 2; Blue color: rescaled-attention weights = 0. (b) Patch clusters on the WSI level and example patches from each cluster. No patch belongs to cluster 4 for this chosen WSI.

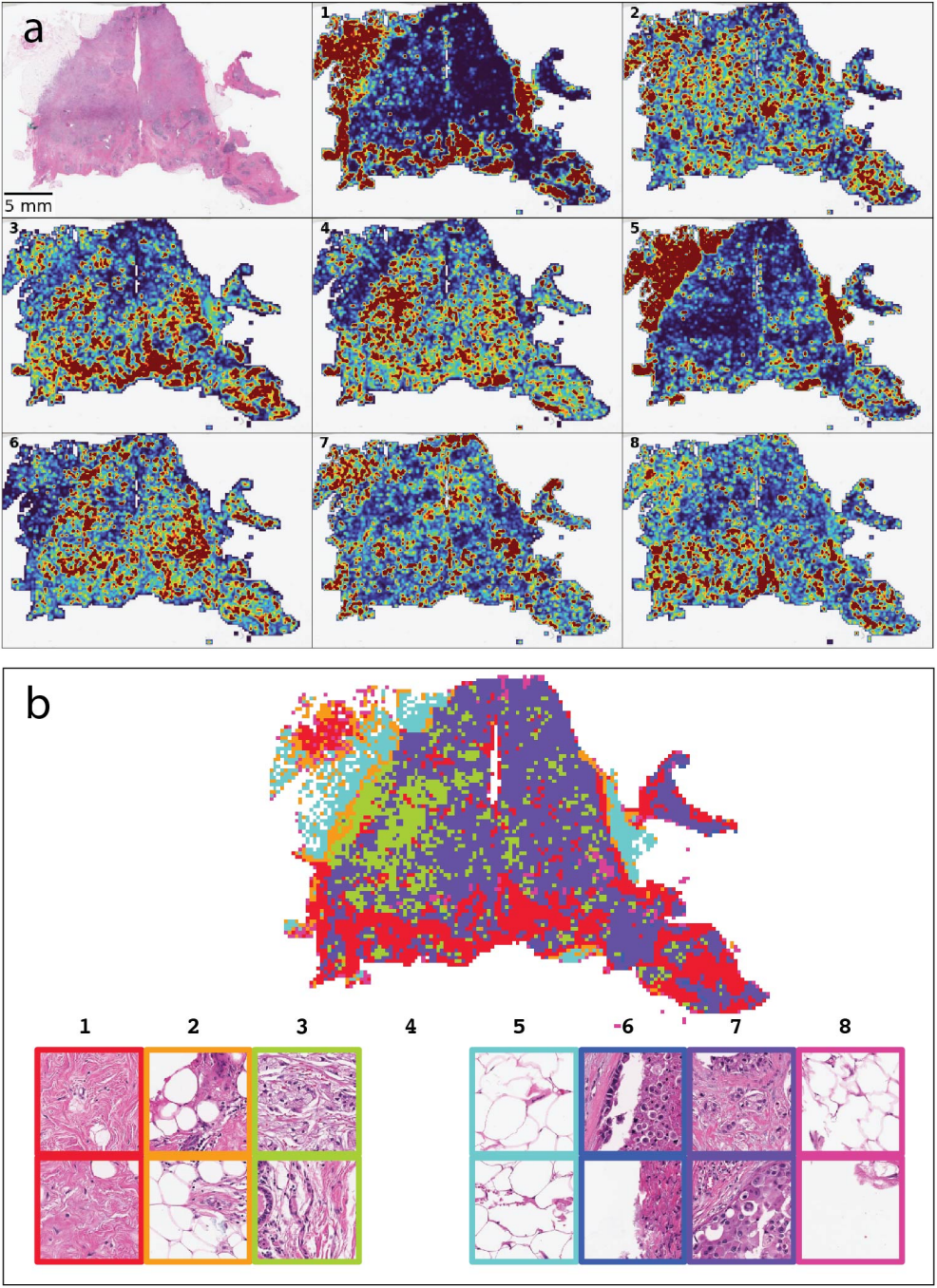

Supplementary Figure S4. Visualization of head-wise attention map and patch clusters for one sample WSI from COAD. (a) Head-wise attention map. Red color: rescaled-attention weights > 2; Blue color: rescaled-attention weights = 0. (b) Patch clusters on the WSI level and example patches from each cluster.

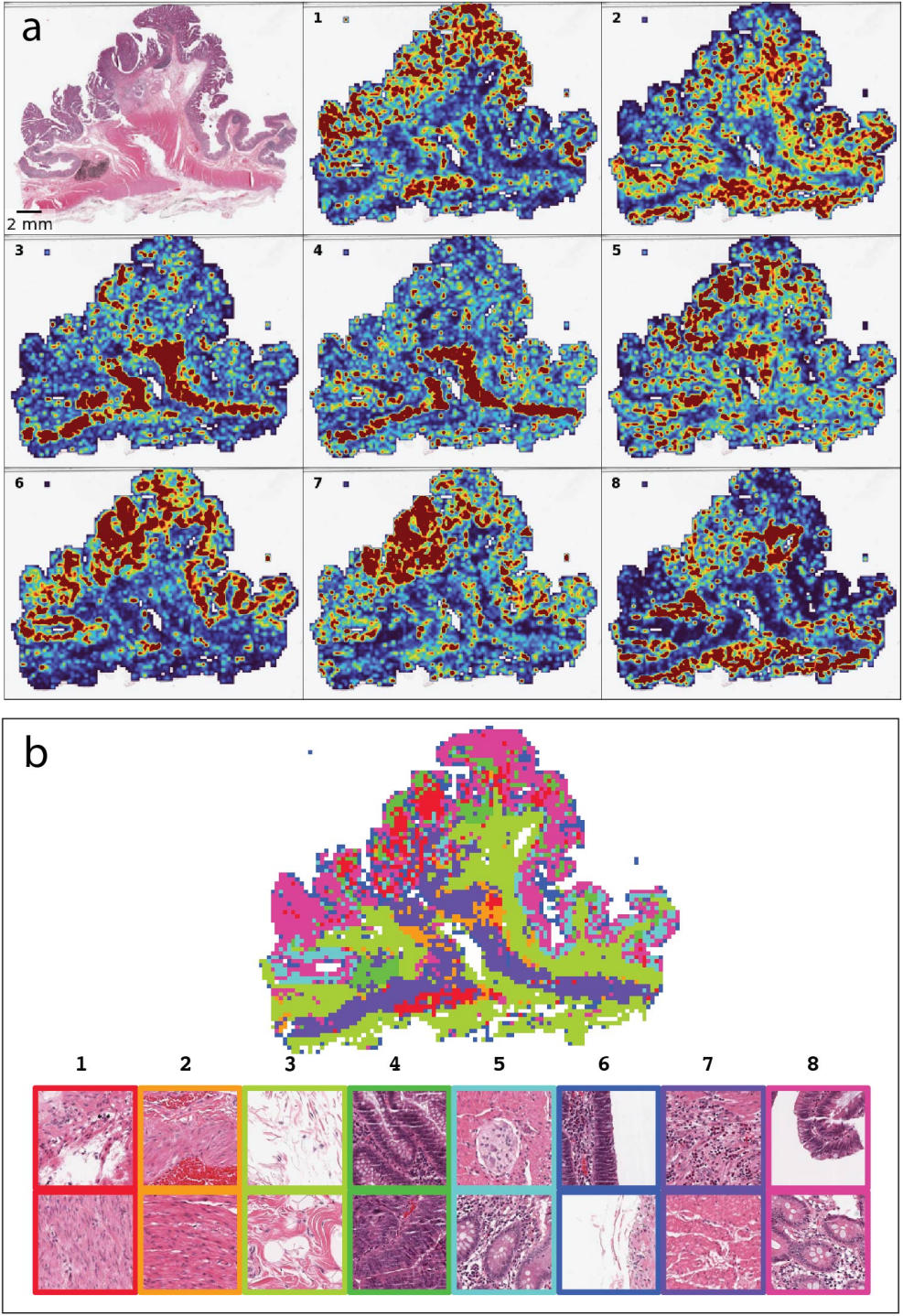